# Shattering the glass ceiling? How the institutional context mitigates the gender gap in entrepreneurship

Christopher J. Boudreaux
Florida Atlantic University
cboudreaux@fau.edu

Boris Nikolaev
Baylor University
Boris_Nikolaev@baylor.edu

**Abstract**

We examine how the institutional context affects the relationship between gender and opportunity entrepreneurship. To do this, we develop a multi-level model that connects feminist theory at the micro-level to institutional theory at the macro-level. It is hypothesized that the gender gap in opportunity entrepreneurship is more pronounced in low-quality institutional contexts and less pronounced in high-quality institutional contexts. Using data from the Global Entrepreneurship Monitor (GEM) and regulation data from the economic freedom of the world index (EFW), we test our predictions and find evidence in support of our model. Our findings suggest that, while there is a gender gap in entrepreneurship, these disparities are reduced as the quality of the institutional context improves.

## 1. Introduction

Why are there gender differences in the performance of high growth ventures? Some argue that women have less start-up capital, human capital, and work experience than their male counterparts (Alsos, Isaksen, & Ljunggren, 2006; Fairlie & Robb, 2009; Fischer, Reuber, & Dyke, 1993). Others say that women have different priorities than men, i.e. ‘social feminist theory’ (Fischer et al., 1993) or often choose entrepreneurship for reasons such as flexibility or to circumvent the ‘glass ceiling’ in traditional employment settings (Fairlie & Robb, 2009; Kephart & Schumacher, 2005; Shane, 2008). Recent advances, however, have shed light on these gender disparities, and argue that once gender selection characteristics have been sufficiently considered, e.g. firm size, sector, and risk preferences, there is no difference between genders when it comes to entrepreneurial performance (Orser, Riding, & Manley, 2006; Robb & Watson, 2012).

Despite these advances, gender gaps in entrepreneurial activity still persist (Hechavarría, Terjesen, Stenholm, Brännback, & Lång, 2017). More work needs to be done to explain differences in entrepreneurship, especially as it relates to high growth entrepreneurship. Thus, the purpose of this study is to examine which factors help mitigate gender differences in opportunity entrepreneurship. We propose that one factor—the institutional context—is a vital component that needs to be considered when examining why the gender gap in entrepreneurship remains. We measure the quality of the institutional context by examining the quality of the regulatory environment. There are several reasons to expect the regulatory environment to effect the gender gap. Financial regulations affect the availability of capital. Good financial regulations ease liquidity constraints, which might disproportionately harm female entrepreneurs. Business regulations affect the ease of doing business, which might also have important gender effects, such as the role of social capital and how it disproportionately helps male entrepreneurs. Lastly, labor regulations might also play an important role, if one considers how rampant sexual violence has affected women in the workplace. Importantly, we hypothesize that gender differences in entrepreneurship are less pronounced as the quality of the institutional context improves.

Using data from 34 countries available in the Global Entrepreneurship Monitor (GEM) and regulatory data from the economic freedom of the world index (Gwartney, Lawson, & Hall, 2016), we find that female entrepreneurs are less involved in opportunity entrepreneurship than their male counterparts. Once one considers the quality of the regulatory environment, however, we find that the gender differences are reduced. More specifically, our results indicate that women are just as likely to be involved in opportunity entrepreneurship when burdensome regulations have been removed.

These findings are important for several reasons. First, our findings examine important gender differences in opportunity entrepreneurship. While our results do reveal these gender differences, our findings suggest these differences can be mitigated. Because we find little difference in opportunity entrepreneurship between genders in a high quality regulatory environment, we are hopeful that our findings will increase the attention paid to how regulations affect gender. Second, and relatedly, these findings have important policy implications. If one desires to remove gender differences and offer an equal playing field, then our study suggests that revisiting the regulatory environment is a good place to start. Recently, workplace sexual violence has been shed to light, which has brought these gender differences to the forefront of contemporary issues. Policymakers might look to begin examining how regulations have difference effects on men and women and seek to improve these gender differences by improving the quality of the existing regulatory environment.

## 2. Theory and hypotheses

*2.1. Feminist theory*

Before we proceed with our theory, it is important to clarify what is meant by 'feminist theory'. As explained in (Watson, 2002, p. 91), there are two types of feminist theory:

> "*Liberal feminist theory* (Fischer et al., 1993) suggests that small and medium enterprises (SMEs) run by women will exhibit poorer performance because women are overtly discriminated against (by lenders, for example) or because of other systematic factors that deprive women of important resources (for example, business education and experience). By way of contrast, *social feminist theory* (Fischer et al., 1993) suggests that men and women are inherently different by nature. These differences do not imply that women will be less effective in business than men, but only that they may adopt different approaches, which may or may not be as effective as the approaches adopted by men."

In this study, we argue that regulatory barriers might play an important role in shaping the gender gap in entrepreneurship, which is based on *liberal feminist theory* (Fischer et al., 1993). Certainly, there might be some merit to examining socio-cultural approaches consistent with *social feminist theory*, but we believe this is less relevant in our context.

Feminist theory[1] recognizes that there are inherent cultural and gender biases toward feminine characteristics and especially women. Aristotle, for instance stated that "women were weak, cautious, domesticated, and nurturing while men occupy the opposite stance thus, making them naturally superior" (Marlow & Patton, 2005, p. 720). Further, men have been equated with "the male, public citizen who is deemed rational, abstract, impartial, independent, active, and strong whereas women, linked with the private sphere of the home, are characterized as noncitizens as they are assumed to be emotional, irrational, dependent, passive, and focused upon domestic concerns" (Lister, 2003, p. 71).

[1] This section draws heavily from (Marlow & Patton, 2005).

These socio-cultural biases are problematic if one considers how society often devalues female credibility. Occupational segregation and domestic/caring responsibilities are prevalent for women (Maushart, 2008), and these "splits" (Hall, 1997) often act as impediments for women to acquire credibility and raise capital (Marlow, 2002). Hence, in the context of entrepreneurship, feminist theory suggests that women face substantial hurdles in the venture creation process. The liberal feminist solution[2] is to remove the financial, administrative, and labor market barriers that disproportionately affect women, and consequently, "level the playing field" (Cockburn, 1991; Marlow & Patton, 2005). Accordingly, we tie liberal feminist theory to institutional theory to relate how these regulatory barriers might disproportionately affect women.

### *2.2. Institutional theory*

Institutions are "the humanly devised constraints that shape human interaction" (North, 1990, p.3). Institutions "consist of both informal constraints (sanctions, taboos, customs, traditions, codes of conduct), and formal rules (constitutions, laws, property rights)" (North, 1990). Informal constraints refer to the norms of social customs and are often referred to as "culture" whereas formal rules are created by the government and represent our laws we must abide by.

Williamson (2000) illustrates these formal rules and informal sanctions using a conceptual framework in a four-level hierarchy, which has been recently applied to the institutional context of entrepreneurial action (Bylund & McCaffrey, 2017; Estrin, Korosteleva, & Mickiewicz, 2013;

[2] It is important to mention that social feminist theory is critical of such a solution. Social feminist theory argues that these biases are endemic in society and culture, and as such, removing regulatory barriers will not improve the inherent biases that affect women (Marlow & Patton, 2005).

Misangyi, Weaver, & Elms, 2008; Pacheco, York, Dean, & Sarasvathy, 2010). This framework begins at the top (level one) with the informal constraints (i.e., customs, traditions, and norms). These institutions are entrenched in society and emerge spontaneously over a long period of time (100 to 1000 years). Formal institutions (level two) represent the institutional environment that defines the "rules of the game" (North, 1991 p.98), which take less time to change (10 to 100 years). Formal institutions are the rules that define action, which often define property rights and regulatory actions. These are the rules that entrepreneurs must abide by. Governance (level three) represents the play of the game or how governance structures align with transactions. Governance structures take even less time to change (1 to 10 years). Lastly, individual action (level four) depicts the choices individuals make, which includes resource allocation and employment choices (e.g. entrepreneurship). The choices entrepreneurs make depend critically on the higher three levels of hierarchy (Bylund & McCaffrey, 2017; Williamson, 2000).

Institutions are considered vital for entrepreneurship (Acemoglu, Johnson, & Robinson, 2005; Baumol, 1990; Williamson, 2000), and evidence suggests that high quality pro-market institutions encourage productive entrepreneurship and innovation (Bjørnskov & Foss, 2010a, 2016; Boudreaux, 2014, 2017; Boudreaux, Nikolaev, & Klein, 2018; McMullen, Bagby, & Palich, 2008; Nyström, 2008; Sobel, 2008; Tebaldi & Elmslie, 2013). Depending on the context, institutions can either encourage innovation and the market process (i.e. productive), encourage redistributive effects (i.e. unproductive), or encourage rent seeking and the creation of entry barriers (i.e. destructive) to reduce contestability (Baumol, 1990; Sobel, 2008). This occurs because the institutional context affects the *allocation* of the supply of entrepreneurs towards different sectors (Boudreaux, Nikolaev, & Holcombe, 2017; Gohmann, Hobbs, & McCrickard, 2008; Murphy, Shleifer, & Vishny, 1991). When the returns to productive entrepreneurship exceed the returns to lobbying, entrepreneurs find it more profitable to engage in productive entrepreneurship and vice versa. In support of these findings, recent sensitivity analyses conclude that economic institutions (level two) are the strongest antecedents of opportunity-motivated entrepreneurship across countries (Nikolaev, Boudreaux, & Palich, 2018), which supports a general consensus that government size, the tax burden, and the welfare state are robustly negatively correlated with entrepreneurial activity (Bjørnskov & Foss, 2010b; Boudreaux, Nikolaev, & Klein, 2017; Lihn & Bjørnskov, 2017; Nyström, 2008).

### *2.3. Hypotheses development*

#### *2.3.1. Financial regulations*

A substantial literature indicates that financial capital[3] is an important antecedent of entrepreneurship (Acs & Szerb, 2007; Fairlie & Krashinsky, 2012). By alleviating liquidity constraints[4], financial capital helps assist nascent firm survival (Blanchflower & Oswald, 1998; Evans & Jovanovic, 1989; Holtz-Eakin, Joulfaian, & Rosen, 1994; Lindh & Ohlsson, 1996)—especially during firms' formative years (Bates, 1990). However, there is reason to believe that one's gender plays an important role in the credit rationing decision, and this might consequently explain some of the variation in entrepreneurial activity between genders.

Liberal feminist theory (Fischer et al., 1993) argues that women face discrimination in financial lending. Because women face socio-cultural biases (Elizabeth & baines, 1998; Minniti & Nardone, 2007), women are perceived to be less credible than men (Marlow & Patton, 2005).

[3] Financial capital is measured as household income, which is strongly correlated with wealth (Bricker, Henriques, Krimmel, & Sabelhaus, 2016; Saez & Zucman, 2016).

[4] Hurst & Lusardi (2004) argue that liquidity constraints are not really present as the majority of the relationship between assets and entrepreneurial entry is found only for those with wealth beyond the 95th percentile in the wealth distribution. However, (Fairlie & Krashinsky, 2012) bifurcate samples into opportunity and necessity entrepreneurs and finds that, when this selection bias is considered, liquidity constraints are found to be present.

For instance, findings from the Diana Project indicate that women face gender myths, which hinder their ability to raise venture capital (Brush, Carter, Gatewood, Greene, & Hart, 2008). Studies suggest that men have better access to capital than women—especially external equity capital (Orser et al., 2006)—and women must pay higher interest rates, on average, when they do gain access to the loans (Muravyev, Talavera, & Schäfer, 2009). This has important implications for entrepreneurship.

We expect it is more difficult for women to form opportunity-motivated entrepreneurial ventures in low-quality financial regulatory environments. In this conext, entrepreneurs must rely on their own sources of capital and funding because it is more difficult to navigate the difficult financial regulations. This is especially true for women, who face discrimination in lending (Muravyev et al., 2009). Burdensome financial regulations deter entry and existing job growth by increasing the administrative burden (Djankov, La Porta, Lopez-de-Silanes, & Shleifer, 2002). This is important because venture capital has been shown to increase innovation (Kortum & Lerner, 2001).

In contrast, it should become easier for women to establish opportunity-motivated entrepreneurial ventures in high-quality financial regulatory environments. An environment that has high-quality financial regulations has lower interest rates and better access to capital and other external sources of funding, such as collateral (Simoes, Crespo, & Moreira, 2016). Therefore, increasing the quality of financial regulations should make it easier for women to receive financial capital. Consequently, this should help alleviate gender disparities in lending, which should provide a more even playing field for women, since women now find it easier to access capital. Finally, if women find it easier to access capital, in high-quality financial regulatory environments, then men will no longer have an advantage over women, all else held equal. For these reasons, we propose our first hypothesis:

**Hypothesis 1**. Women are less involved in opportunity entrepreneurship than men, but improving the quality of credit market regulations reduces this gender gap.

*2.3.2. Business regulations*

High entry barriers reduce new venture entry (Dean & Meyer, 1996), and one such entry barrier—business regulations—has been shown to deter new venture start-up rates (De Soto, 2000; Djankov et al., 2002) and growth rates of high growth ventures. These business regulations increase the costs of doing business including licensing restrictions, administrative requirements, bureaucracy costs, tax compliance, and even the costs associated with bribes and favoritism (Djankov et al., 2002; Gwartney et al., 2016). Business regulations deter entry by increasing the costs of new venture formation (Ho & Wong, 2007), however, there are reasons to believe that business regulations might have different effects on opportunity entrepreneurship for female entrepreneurs than male entrepreneurs.

Liberal feminist theory (Fischer et al., 1993) argues that women face discrimination that is imbued in socio-cultural biases. These biases, in turn, deter entry and existing firm growth through high regulatory costs. These burdens have different effects on women than men. We expect that it is more difficult for women to form productive entrepreneurial ventures in low-quality administrative regulatory environments that have higher costs of doing business. For instance, studies show that highly regulated economies are susceptible to corruption (Holcombe & Boudreaux, 2015). If bribes are a cost of doing business that is often required to get the business established (De Soto, 2000), then it is reasonable to believe that these highly regulated environments might disproportionately harm women because women bribe less than men (Swamy, Knack, Lee, & Azfar, 2001).

However, it should be easier for women to form productive entrepreneurial ventures in a high-quality administrative regulatory environment. In contrast to the low-quality environments, there is less corruption (Mauro, 1995; Mo, 2001), and because women bribe less than men (Swamy et al., 2001), women are less affected by these burdensome administrative and regulatory costs. Thus, the high quality business regulatory environment should even the playing field, which is important because it can help reduce discrimination, which has been argued to attribute to gender disparities in areas such as lending and consumption (Fairlie & Robb, 2009). Based on these findings we propose that:

**Hypothesis 2**. Women are less involved in opportunity entrepreneurship than men, but improving the quality of business regulations reduces this gender gap.

*2.3.3. Labor market regulations*

Liberal feminist theory (Fischer et al., 1993) argues that women face discrimination in society. While we have suggested this discrimination affects financial and business regulations, it is also possible that the discrimination works through labor regulations, such as hiring and firing regulations, the costs of worker dismissals, and collective bargaining issues.

An important literature on labor market regulations and entrepreneurial intensity explains how restrictive labor market regulations reduce entrepreneurship rates across countries (Acs & Szerb, 2007; Van Stel, Storey, & Thurik, 2007). Consider, for example, how more flexible labor regulations might influence entrepreneurial activity:

> "On the side of employees, the safety of their paid job is less which may make them more likely to decide to start their own business (push effect). On the side of the entrepreneurs, they have more flexibility in running their business which makes business ownership more attractive (pull effect)" (Stel et al., 2007, p. 182).

Moreover, there are reasons to believe that these labor market regulations might have important gender effects on the decision to innovation and create jobs.

We expect that labor market regulations disproportionately affect the propensity of women to engage in productive forms of entrepreneurship in low quality labor regulatory environments. In these environments, labor regulations are rigid. It is difficult to fire workers, which reduces the incentive for entrepreneurs to seek self-employment or entrepreneurship. This is especially true for women who face gender biases such as occupational segregation (Lerner, Brush, & Hisrich, 1997) and gender-based occupational stereotypes (Eccles, 1994; Eccles, Wigfield, Harold, & Blumenfeld, 1993). If women face biases in entrepreneurship, and labor markets are more rigid, then low-quality labor regulatory environments provide fewer incentives for women to quit traditional employment in the hopes of starting a new venture.

In contrast, women's propensity to form productive entrepreneurial ventures is less affected in high quality labor regulatory environments because these environments promote job flexibility. When the labor market is more flexible, entrepreneurs can run their business in more attractive ways. They have more freedom to hire and fire workers and are not penalized for this flexibility (Gwartney et al., 2016). This is consistent with findings that economic freedom correlated with higher women's rights (Fike, 2017). In these environments, women have greater incentives to be creative and innovate. Similarly, more job flexibility is associated with higher rates of new firm entry (Van Stel et al., 2007), which might equal the playing field between genders. Thus, we propose that:

**Hypothesis 3**. Women are less involved in opportunity entrepreneurship than men, but improving the quality of labor market regulations reduces this gender gap.

## 3. Data and analysis

### *3.1. Dependent variable*

Our dependent variable of interest in this study is taken from the Global Entrepreneurship Monitor (GEM) (Reynolds et al., 2005). Opportunity-motivated entrepreneurship, is taken from the GEM variable, TEAYYOPP. This variable asks whether an individual is involved in total early stage entrepreneurial activity (TEA). If the answer is yes, it then asks whether the reason for the involvement was due to a perceived opportunity or out of necessity. We use this variable to create our measure of opportunity entrepreneurship, which takes a value of 1 if an individual is involved in early stage entrepreneurial activity to capture an opportunity and 0 otherwise.

### *3.2. Predictor variables*

We use regulatory data from the economic freedom of the world index (Gwartney et al., 2016) to construct our regulatory measure. Refer to the appendix for a more complete description of these measures and their sources. Regulation is the fifth area component of the economic freedom of the world index (EFW), and it is comprised of three sub-components including (a) credit market regulations, (b) labor market regulations, and (c) business regulations. Our analysis involves an examination of the summary measure of regulation (area five of EFW) as well as each of these area five subcomponents. Credit market regulations are calculated as the average of three measures including (i) ownership of banks, (ii) private sector credit, and (iii) interest rate controls. EFW uses data primarily from the World Bank to compile these capital market measures, which are considered more free with higher proportions of private ownerships of banks, private sector credit, and interest rates determined by market forces. Labor market regulations are calculated as the average of six measures including (i) hiring regulations and minimum wage, (ii) hiring and firing regulations, (iii) centralized collective bargaining, (iv) hours regulations, (v) mandated cost of worker dismissal, and (vi) conscription. EFW uses data primarily from the World Bank doing business report and the World Economic Forum, Global Competitiveness Report to compile these labor market measures. These measures are considered more free when the labor regulations are more flexible, the costs of firing is lower, and conscription is lower. Business regulations are calculated as the average of six measures including (i) administrative requirements, (ii) bureaucracy costs, (iii) starting a business, (iv) extra payments/bribes/favoritism, (v) licensing restrictions, and (vi) cost of tax compliance. EFW uses data primarily from the World Bank doing business report and the World Economic Forum global competitiveness report to compile these business regulation measures. Data on bureaucracy costs, however, are compiled from the regulatory burden risk ratings from the IHS Markit. Capital market regulations, labor regulations, and business regulations are all measured on a scale from 0 to 10, where 10 indicates more free and 0 indicates less free.

We also include a measure for the entrepreneur's gender. This variable is dummy coded 1 if the entrepreneur is female and 0 if male. Gender data are taken from the Global Entrepreneurship Monitor (Reynolds et al., 2005).

### *3.3. Control variables*

In addition to our predictor variables of regulation and gender, we also include other individual-level variables that have been shown to correlate with entrepreneurship. Age and Age (squared) are continuous variables that denote the age of the entrepreneur and its squared value, respectively. We include an entrepreneur's age and its squared value to be consistent with prior studies on the aging entrepreneur (Kautonen, Kibler, & Minniti, 2017; Lévesque & Minniti, 2006) as well as others that control for curvilinear effects (Wennberg, Pathak, & Autio, 2013). High school education is measured as whether an individual has at least graduated from high school or its equivalent (secondary education) or not. It is calculated from the GEMEDUC harmonized education variable where it takes a value of 1 if an individual has a high school education and 0 otherwise. Household income is taken from the variable, GEMHHINC, which is measured in income terciles. Household income is coded 1 if an individual's household income is in the highest income tercile and 0 if it is in the middle or lowest tercile. Entrepreneurial ties is a measure of an individual's ties with other entrepreneurs. Entrepreneurial ties is coded 1 if an individual knows someone who has created a business in the past two years and 0 otherwise. Self-efficacy is coded 1 if the individual entrepreneur believes he or she has the knowledge, skills, and experience required to start a new business and 0 otherwise. Opportunity recognition is coded 1 if the entrepreneur envisions good business opportunities in the next six months and 0 otherwise. Fear of failure is coded 1 if the entrepreneur responds that fear of failure is likely to prevent him or her from starting a business and 0 otherwise. Recent research supports their importance in predicting and modifying entrepreneurial activity (Boudreaux & Nikolaev, 2018; De Clercq, Lim, & Oh, 2013). These variables are all taken from the Global Entrepreneurship Monitor dataset for the years 2002 to 2012 (Reynolds et al., 2005). We also include a country-level measure of gender equality. This measure is taken from the World Economic Forum for the years 2002 to 2010. This variable is measured on a continuous scale from 0 to 1 where 1 indicates complete equality and 0 indicates complete inequality.

Lastly, we also include control variables at the country-level that are expected to influence entrepreneurial behavior. Log GDP is the natural logarithm of a country's gross domestic product per capita. Log pop is the natural logarithm of a country's total population. These variables are taken from the World Bank's country indicator's database for the years 2002 to 2012. Log GDP is used to control for the 'natural rate' of entrepreneurship in economic development (Wennekers, Wennekers, Thurik, & Reynolds, 2005). Summary statistics and a correlation matrix are presented in Table 1.

===================

Insert Table 1 About Here

===================

*3.4. Estimation methods*

We examine how individual-level relationships between gender and innovation and job creation are moderated by country-level measures of regulation. Due to the multi-level nature of these relationships, we must control for the fact that standard estimation techniques (e.g. OLS) in the presence of clustered data increases the possibility of Type 1 errors. Research indicates that such standard errors are underestimated due to their non-normal distribution (Hofmann, Griffin, & Gavin, 2000). In our multilevel models, random effects denote country-specific effects that are assumed to effect innovation and job creation, and their use assumes that the groups are drawn randomly from a larger population (Autio, Pathak, & Wennberg, 2013; Peterson, Arregle, & Martin, 2012).

To estimate the influence of country-level factors on an individual's likelihood of participating in opportunity entrepreneurship (binary coded), we use a multi-level logistic regression model that assumes unobserved country-specific effects ($\boldsymbol{u_i}$) to be randomly distributed with a mean of zero, constant variance ($\boldsymbol{u_i} \approx \mathbf{IID}\ (\mathbf{0}, \boldsymbol{\sigma_u^2}\ )$), and uncorrelated to the predictor variables. This estimator permits the intercept and standard errors to vary randomly across countries (Raudenbush, 1988), and provides greater weights to groups with more reliable level 1 estimates, which in turn, provide greater influence in the level 2 regression (Hofmann et al., 2000).

## 4. Results

We begin our analysis with an examination of gender differences in opportunity entrepreneurship. We than proceed to analyze how gender differences in opportunity entrepreneurship are moderated by the quality of the regulatory environment. These results are presented in Table 3. More specifically, model 1 is our baseline model that includes a gender dummy, a measure of the quality of the regulatory environment, and a vector of control variables. Model 2 augments this model to include the interaction between the regulatory variable and the gender dummy. Models 3-5 repeat this analysis but replace the summary regulatory measure with its component areas. These areas include: (a) labor market regulations, (b) credit market regulations, and (c) business regulations.

===================
Insert Table 2 About Here
===================

The results from model 1 indicate that there are significant gender differences in opportunity entrepreneurship. Across all models, we find a negative and statistically significant effect of gender (female) on the propensity to be engaged in opportunity entrepreneurship, which is consistent with claims that a gender gap persists in entrepreneurial activity (Fairlie & Robb, 2009; Hechavarría et al., 2017). More importantly, the findings from the interaction models suggest that the quality of the regulatory environment plays an important role in explaining the entrepreneurial gender gap. While female entrepreneurs are less involved in opportunity entrepreneurship, the quality of the regulatory environment reduces this gender gap. This can be observed in all moderating effects except for business regulations. These findings support hypotheses 1 and 3 but fails to support hypothesis 2. Based on these results, we conclude that gender differences in entrepreneurial activity are largest when there are burdensome financial and labor regulations but not business regulations. However, as the quality of these regulatory environments improve, we find reductions in the size of the gender gap.

=========================
Insert Figures 1, 2, and 3 About Here
=========================

Because interaction effects are notoriously difficult to interpret in non-linear models such as logit (Ai & Norton, 2003), we plot the interaction effects in Figures 1, 2, and 3. The findings from these interaction effects support the empirical analysis in Table 2. Consistent with the findings in Table 2, we find no evidence that the quality of the business regulations moderates gender differences in opportunity entrepreneurship, and surprisingly, the quality of business regulations is associated with reductions in entrepreneurial activity.

## 5. Discussion, limitations, and concluding remarks

### *5.1. Discussion*

Our results indicate that the relationship between gender and business performance depends on the quality of the regulatory environment. While we find that women generally are less involved in opportunity entrepreneurship when compared to men, these gender differences are reduced as the quality of the regulatory environment improves. We used insights from feminism theory and institutional theory to hypothesize that women are more likely to face discrimination in low-quality regulatory environments, which explains why gender differences are most prominent in these environments.

These findings have important implications. If policy makers desire to reduce the gender gap in business performance, our evidence suggests that policies designed to enhance the quality of the regulatory environment is a good place to start.

### 5.2. Limitations

Inevitably, our study does face some limitations. Our findings are not unanimous for all three regulatory measures. While our findings largely suggest that gender differences in opportunity entrepreneurship are most pronounced in low-quality regulatory environments and become less pronounced as the quality of the regulatory environment improves, we do not find any evidence to suggest that gender differences in opportunity entrepreneurship on the quality of business regulations. Therefore, future work should emphasize why different regulations might have different effects on the gender gap in entrepreneurial activity.

### 5.3. Concluding remarks

We hypothesized that gender differences in opportunity entrepreneurship depends on the quality of the regulatory environment. We used economic freedom (Gwartney et al., 2016) to measure the quality of the regulatory environment, and we found that, with the exception of business regulations, gender differences in entrepreneurial activity are most pronounced in low-quality regulatory environments, but these differences are less pronounced as the quality of the regulatory environment improves. More specifically, women are less likely to be involved in opportunity entrepreneurship than men at low levels of capital market regulations and labor market regulations, but this difference decrease as the quality of these regulations increases.

These findings are important because our results suggest that enhancing the quality of the regulatory environment would help to reduce gender differences in business performance. If policy makers desire to reduce gender disparities in entrepreneurial activity, reducing credit market constraints and labor market constraints are good places to start with addressing gender disparities.

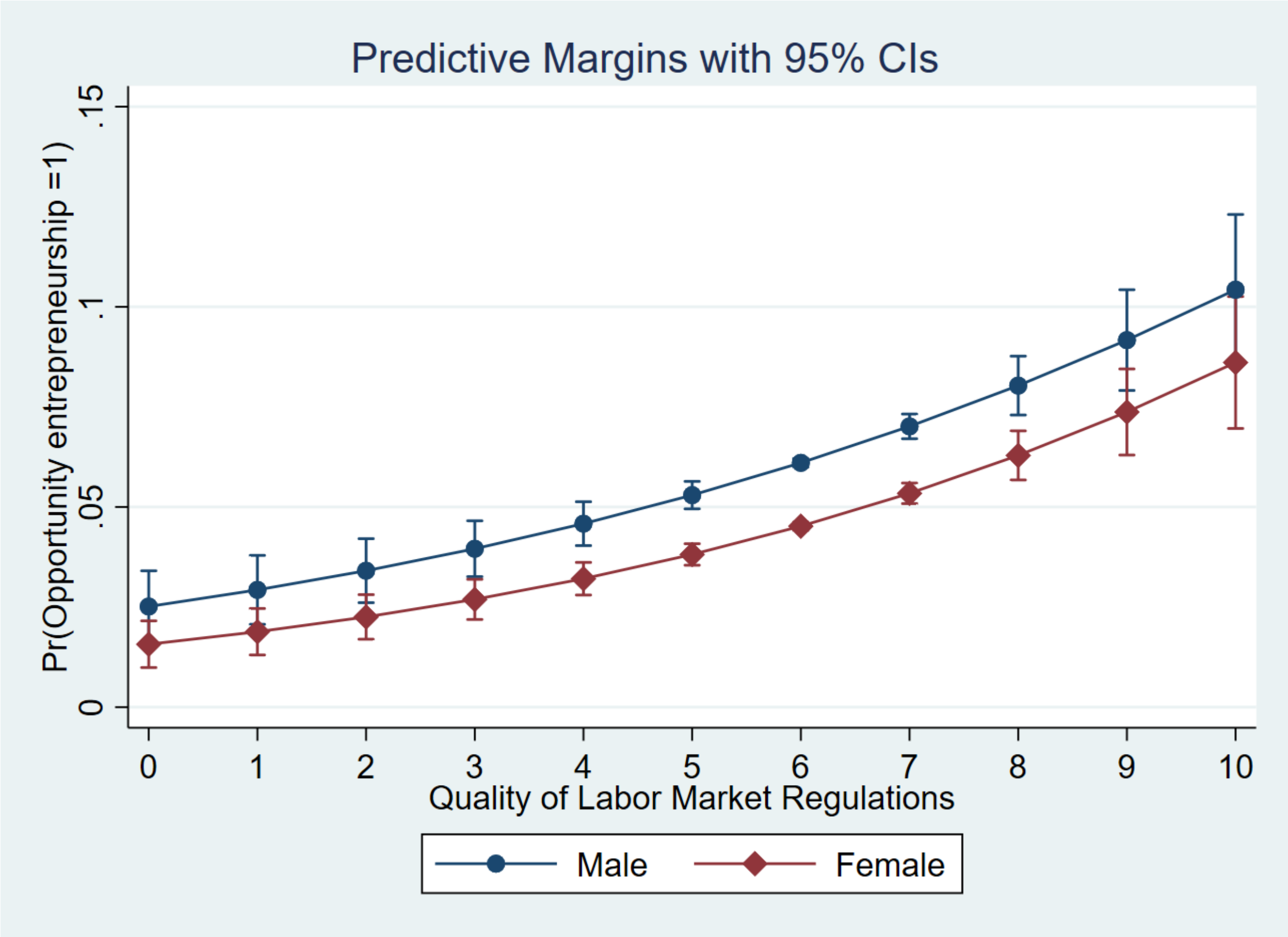

**Figure 1.** Interaction between labor market regulations and gender on opportunity entrepreneurship

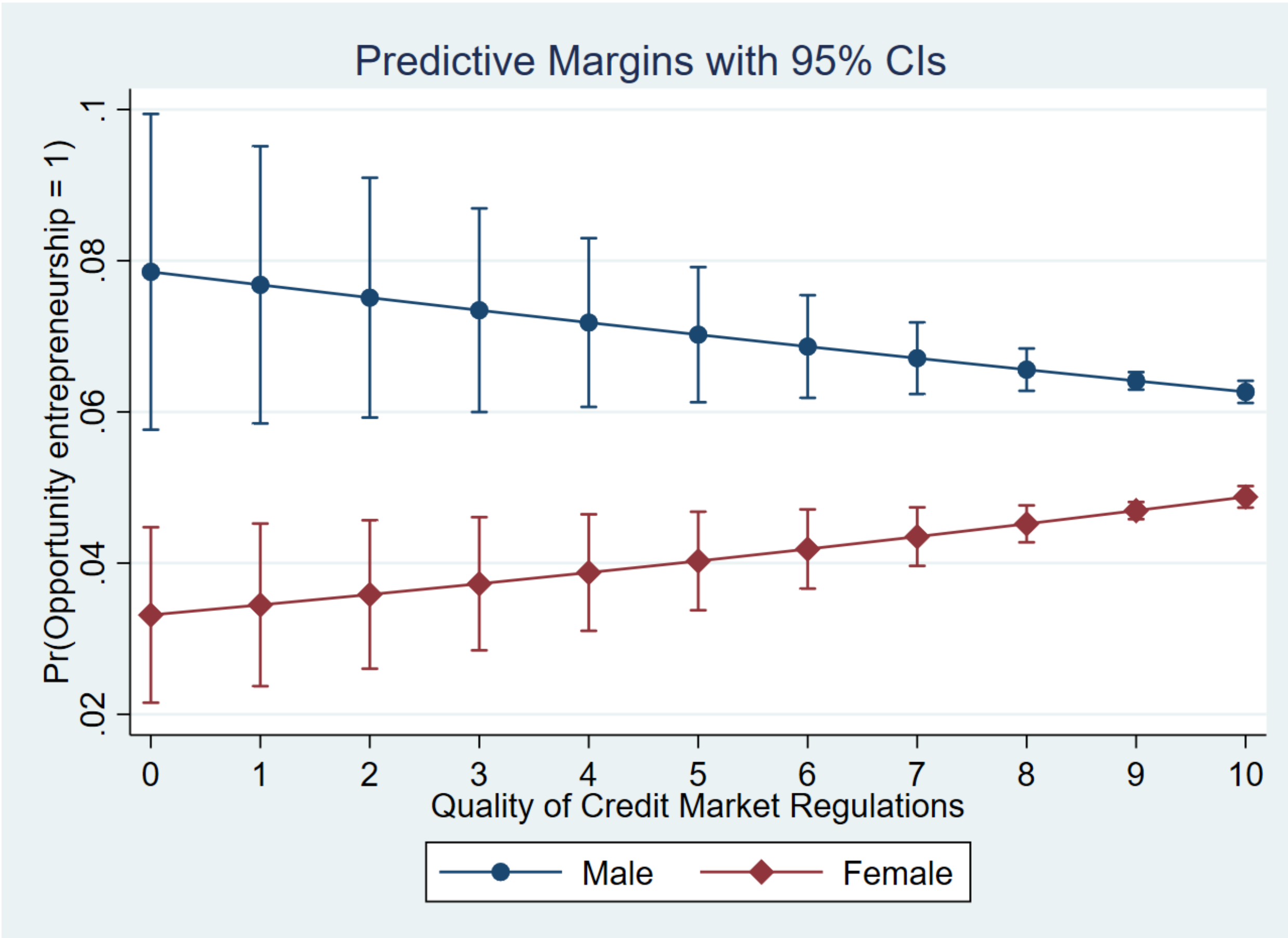

**Figure 2.** Interaction between credit market regulations and gender on opportunity entrepreneurship

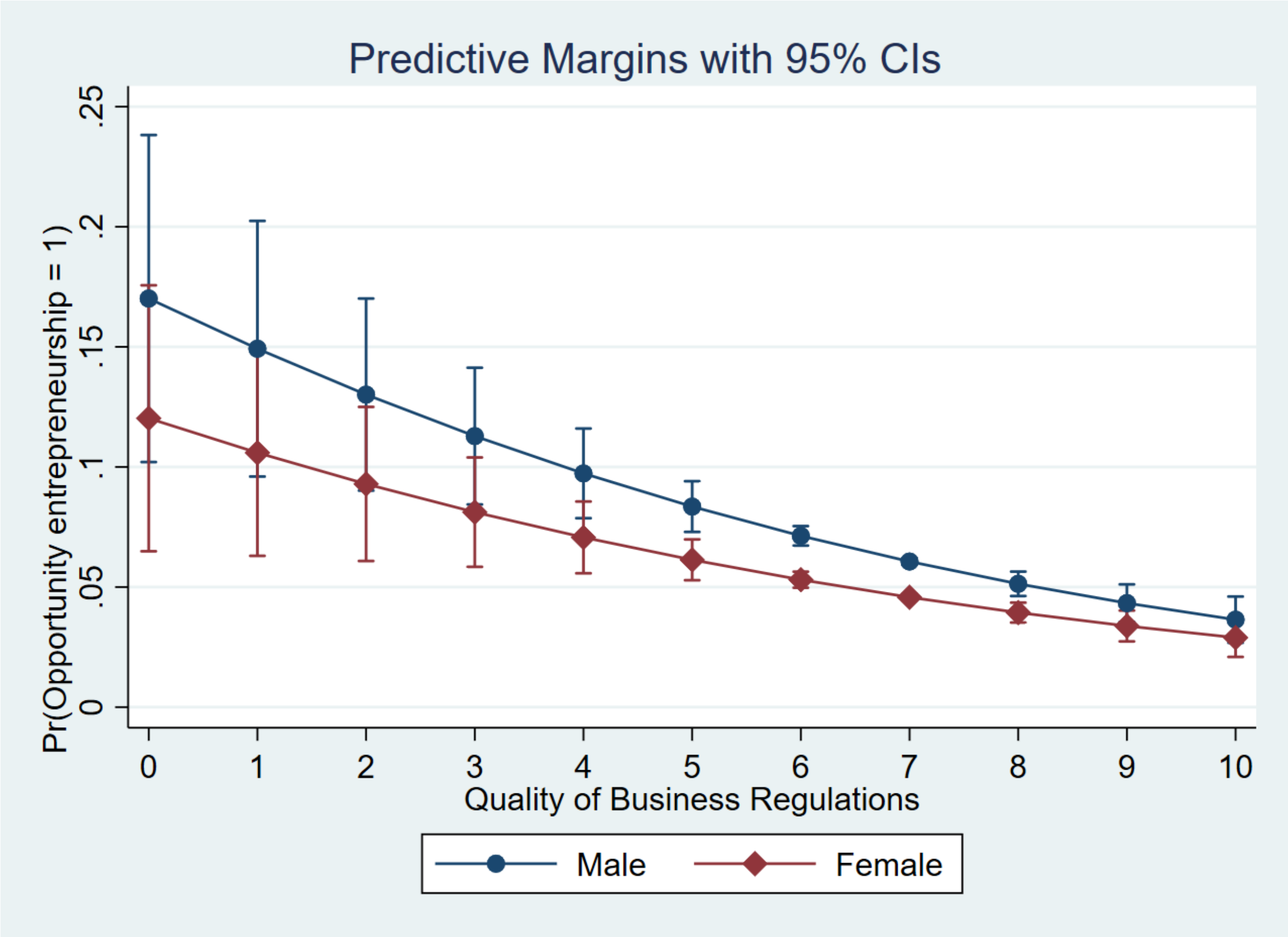

**Figure 3.** Interaction between business regulations and gender on opportunity entrepreneurship

**Table 1.** Summary statistics and correlation matrix

| Variable | Mean | Std dev | | **1** | **2** | **3** | **4** | **5** | **6** | **7** | **8** | **9** | **10** | **11** | **12** | **13** | **14** | **15** | **16** | **17** |
|---|---|---|---|---|---|---|---|---|---|---|---|---|---|---|---|---|---|---|---|---|
| Opportunity entrepreneurship | 0.057 | 0.231 | **1** | 1 | | | | | | | | | | | | | | | | |
| Age | 43 | 13.9 | **2** | -0.06 | 1.00 | | | | | | | | | | | | | | | |
| Age (squared) | 1999 | 1230.2 | **3** | -0.07 | 0.98 | 1.00 | | | | | | | | | | | | | | |
| Female | 0.511 | 0.50 | **4** | -0.08 | 0.03 | 0.02 | 1.00 | | | | | | | | | | | | | |
| Household income | 0.797 | 0.40 | **5** | 0.03 | -0.05 | -0.05 | -0.05 | 1.00 | | | | | | | | | | | | |
| Education | 0.677 | 0.47 | **6** | 0.05 | -0.12 | -0.14 | -0.01 | 0.06 | 1.00 | | | | | | | | | | | |
| Self-efficacy | 0.475 | 0.50 | **7** | 0.21 | -0.01 | -0.03 | -0.17 | 0.06 | 0.10 | 1.00 | | | | | | | | | | |
| Opportunity recognition | 0.329 | 0.47 | **8** | 0.13 | -0.07 | -0.07 | -0.08 | 0.02 | 0.06 | 0.17 | 1.00 | | | | | | | | | |
| Fear of failure | 0.379 | 0.49 | **9** | -0.08 | -0.03 | -0.04 | 0.07 | -0.02 | -0.04 | -0.13 | -0.06 | 1.00 | | | | | | | | |
| Entrepreneurial ties | 0.366 | 0.48 | **10** | 0.14 | -0.15 | -0.16 | -0.12 | 0.07 | 0.09 | 0.24 | 0.21 | -0.03 | 1.00 | | | | | | | |
| Gross domestic product [a] | 34.4 | 13.4 | **11** | 0.02 | 0.04 | 0.03 | 0.01 | -0.04 | 0.06 | 0.01 | 0.12 | -0.03 | 0.02 | 1.00 | | | | | | |
| Population [a] | 35400 | 26800 | **12** | -0.02 | 0.05 | 0.04 | 0.02 | -0.07 | 0.02 | -0.02 | -0.10 | 0.02 | -0.14 | -0.09 | 1.00 | | | | | |
| Regulation | 7.445 | 0.66 | **13** | 0.02 | 0.06 | 0.06 | 0.04 | -0.06 | 0.09 | -0.02 | 0.11 | -0.08 | -0.04 | 0.51 | 0.09 | 1.00 | | | | |
| Credit market regulations | 9.374 | 0.78 | **14** | 0.00 | 0.03 | 0.02 | 0.00 | -0.02 | -0.08 | 0.01 | 0.08 | 0.01 | -0.01 | 0.29 | -0.01 | 0.44 | 1.00 | | | |
| Labor market regulations | 6.300 | 1.39 | **15** | 0.02 | 0.06 | 0.07 | 0.05 | -0.11 | 0.17 | -0.02 | 0.04 | -0.08 | -0.07 | 0.32 | 0.28 | 0.82 | -0.01 | 1.00 | | |
| Business regulations | 6.666 | 0.73 | **16** | 0.01 | 0.01 | 0.02 | 0.01 | 0.09 | 0.01 | -0.02 | 0.14 | -0.07 | 0.04 | 0.46 | -0.25 | 0.68 | 0.16 | 0.35 | 1.00 | |
| Gender equality | 0.725 | 0.05 | **17** | 0.00 | 0.04 | 0.03 | 0.01 | 0.01 | -0.04 | 0.02 | 0.09 | 0.00 | 0.04 | 0.60 | -0.28 | 0.38 | 0.35 | 0.11 | 0.45 | 1.00 |

***Note.*** [a] Denoted in thousands.

**Table 2.** Effects on Opportunity entrepreneurship

| | Dependent variable = Opportunity Entrepreneurship | | | | | | | | | |
|---|---|---|---|---|---|---|---|---|---|---|
| | (1) | | (2) | | (3) | | (4) | | (5) | |
| Variables | | | | | | | | | | |
| Age | 0.0865*** | (0.00) | 0.0864*** | (0.00) | 0.0866*** | (0.00) | 0.0865*** | (0.00) | 0.0865*** | (0.00) |
| Age (squared) | -0.0013*** | (0.00) | -0.0013*** | (0.00) | -0.0013*** | (0.00) | -0.0013*** | (0.00) | -0.0013*** | (0.00) |
| Female (F) | -0.333*** | (0.01) | -0.351*** | (0.02) | -0.338*** | (0.01) | -0.360*** | (0.02) | -0.336*** | (0.01) |
| Household income | 0.156*** | (0.02) | 0.156*** | (0.02) | 0.161*** | (0.02) | 0.155*** | (0.02) | 0.156*** | (0.02) |
| Education | 0.168*** | (0.02) | 0.168*** | (0.02) | 0.164*** | (0.02) | 0.168*** | (0.02) | 0.169*** | (0.02) |
| Self-efficacy | 1.927*** | (0.02) | 1.928*** | (0.02) | 1.928*** | (0.02) | 1.926*** | (0.02) | 1.927*** | (0.02) |
| Opportunity recognition | 0.606*** | (0.01) | 0.606*** | (0.01) | 0.603*** | (0.01) | 0.606*** | (0.01) | 0.607*** | (0.01) |
| Fear of failure | -0.576*** | (0.02) | -0.575*** | (0.02) | -0.575*** | (0.02) | -0.575*** | (0.02) | -0.575*** | (0.02) |
| Entrepreneurial ties | 0.605*** | (0.01) | 0.605*** | (0.01) | 0.606*** | (0.01) | 0.605*** | (0.01) | 0.606*** | (0.01) |
| GDP (log) PPP | -0.152* | (0.07) | -0.151* | (0.07) | -0.104 | (0.06) | -0.200** | (0.07) | -0.233*** | (0.06) |
| Population (log) | -5.074*** | (0.58) | -5.058*** | (0.58) | -4.350*** | (0.59) | -5.193*** | (0.57) | -5.276*** | (0.57) |
| Gender equality | 1.096* | (0.54) | 1.076* | (0.54) | 0.543 | (0.55) | 1.120* | (0.54) | 1.523** | (0.55) |
| Regulation | | | | | | | | | | |
| Summary measure | 0.0511 | (0.04) | 0.0260 | (0.05) | | | | | | |
| Labor market | | | | | 0.165*** | (0.03) | | | | |
| Credit market | | | | | | | -0.0271 | (0.02) | | |
| Business | | | | | | | | | -0.193*** | (0.05) |
| Moderating hypotheses [a] | | | | | | | | | | |
| F x Summary measure | | | 0.0753*** | (0.02) | | | | | | |
| F x Labor market | | | | | 0.0253* | (0.01) | | | | |
| F x Credit market | | | | | | | 0.0701*** | (0.02) | | |
| F x Business | | | | | | | | | 0.0243 | (0.02) |
| Number of observations | 437559 | | 437559 | | 437559 | | 437559 | | 437559 | |
| Number of groups | 27 | | 27 | | 27 | | 27 | | 27 | |
| AIC | 157606.1 | | 157595.4 | | 157572.5 | | 157595.2 | | 157591.4 | |
| Degrees of freedom | 23 | | 24 | | 24 | | 24 | | 24 | |
| P> chi2 | *** | | *** | | *** | | *** | | *** | |
| Log-likelihood | -78755.1 | | -78748.7 | | -78737.2 | | -78748.6 | | -78746.7 | |
| % of variance, rho | 3.3 | | 3.4 | | 4.4 | | 3.5 | | 4.2 | |
| Variance of random intercept | 0.337 | | 0.337 | | 0.392 | | 0.35 | | 0.38 | |

*Note.* Standard errors in parentheses. [a] Mean-centered. Dependent variable = 1 if involved in opportunity entrepreneurship and 0 otherwise.
* p<0.05
** p<0.01
*** p<0.001